\documentclass{WileyMSP-template}
\usepackage{wrapfig}
\usepackage{amsmath}
\usepackage{hyperref}
\usepackage{cite}

\begin{document}

\pagestyle{fancy}

\title{Enhanced Nonlinearity of Epsilon-Near-Zero Indium Tin Oxide Nanolayers with Tamm Plasmon-Polariton States}

\maketitle


\author{Tornike Shubitidze}
\author{Wesley A. Britton}
\author{Luca Dal Negro*}



\begin{affiliations}
T. Shubitidze\\
Department of Electrical \& Computer Engineering and Photonics Center, Boston University, 8 Saint Mary’s Street, Boston, Massachusetts 02215, USA\\
Dr. W. A. Britton\\
Division of Material Science and Engineering, Boston University, 15 Saint Mary’s Street, Brookline, Massachusetts 02446, USA\\
Prof. L. Dal Negro\\
Department of Electrical \& Computer Engineering and Photonics Center, Boston University, 8 Saint Mary’s Street, Boston, Massachusetts 02215, USA\\
Division of Material Science and Engineering, Boston University, 15 Saint Mary’s Street, Brookline, Massachusetts 02446, USA\\
Department of Physics, Boston University, 590 Commonwealth Avenue, Boston, Massachusetts 02215, USA
Email Address: dalnegro@bu.edu

\end{affiliations}



\begin{abstract}
Recently, materials with vanishingly small permittivity, known as epsilon-near-zero (ENZ) media, emerged as promising candidates to achieve nonlinear optical effects of unprecedented magnitude on a solid-state platform. In particular, the ENZ behavior of Indium Tin Oxide (ITO) thin films resulted in Kerr-type nonlinearity with non-perturbative refractive index variations that are key to developing more efficient Si-compatible devices with sub-wavelength dimensions such as all-optical switchers, modulators, and novel photon detectors. 
In this contribution, we propose and demonstrate enhancement of the nonlinear index variation of 30\,nm-thick ITO nanolayers by silicon dioxide/silicon nitride (SiO$_{2}$/SiN) Tamm plasmon-polariton structures fabricated by radio-frequency magnetron sputtering on transparent substrates under different annealing conditions. In particular, we investigate the linear and nonlinear optical properties of ITO thin films and resonant photonic structures using broadband spectroscopic ellipsometry and intensity-dependent Z-scan nonlinear characterization demonstrating enhancement of optical
nonlinearity with refractive index variations as large as $\Delta{n}\approx{2}$ in the non-perturbative regime. 
Our study reveals that the efficient excitation of strongly confined plasmon-polariton Tamm states substantially boost the nonlinear optical response of ITO nanolayers providing a stepping stone  for the engineering of more efficient infrared devices and nanostructures for a broad range of applications including all-optical data processing, nonlinear spectroscopy, sensing, and novel photodetection modalities.
\end{abstract}


\section{Introduction}
Epsilon-near-zero (ENZ) media are low-refractive index materials that exhibit many fascinating optical properties including a substantial enhancement of nonlinear optical processes such as harmonic generation, frequency mixing and conversion, electro-optical modulation, and light-induced refractive index changes \cite{Liberal2017,Reshef2019_ENZ_Review, Niu2018, Wu2021, Khurgin2021, Ciattoni2016, Guo2016, Bohn2021, Vezzoli2018_Time_Reversal,Capretti2015THG}. In this context, indium tin oxide (ITO) degenerate semiconductors have attracted a great interest due to their ultrafast (sub-picosecond) nonlinear optical responses culminating in the recent demonstrations of order-of-unity light-induced refractive index changes \cite{Alam2016}. These effects, which are boosted in sub-wavelength nanostructures across their ENZ spectral regions, provide novel opportunities for classical and quantum
information technologies, materials analysis, optical spectroscopy, sensing on the scalable silicon photonics platform. Specifically, a number of impressive ENZ-driven optical nonlinear
phenomena and devices have been established based on ITO
nanolayers, including enhanced second-harmonic \cite{Capretti2015SHG} and third-harmonic
frequency generation \cite{Capretti2015THG}, sub-picosecond all-optical modulation \cite{Guo2016,Bohn2021,Gosciniak2023},
high-efficiency optical time reversal \cite{Vezzoli2018_Time_Reversal}, as well as diffraction effects unconstrained by the pump bandwidth for time-varying and spatiotemporal photonic devices \cite{Double_slit,Tirole2022_timeMirror}. Moreover, ENZ structures based on ITO materials feature a
wide tunability of the linear optical dispersion with a zero permittivity
wavelength $\lambda_{ENZ}$ (i.e., the wavelength at which the real part of the permittivity vanishes) that extends from the near-infrared to mid-infrared spectral range \cite{Wang2015,Gui2019}, enabling ENZ-driven nonlinearities across multiple wavelength regimes, only limited by the material optical
losses at the ENZ wavelength. Recently, using magnetron deposition in partnership with X-ray diffraction (XRD), linear and nonlinear optical characterization of ITO thin films, we reported significant enhancement of the nonlinear response of ITO nanolayers through
control of their material microstructure, specifically driven by a secondary phase with anisotropic texturing, which is largely influenced by the deposition and post-deposition vacuum annealing conditions \cite{Britton2022}.

In this work, building on the enhanced optical nonlinearity of ITO thin films with anisotropic crystallographic texturing, we propose and demonstrate Tamm plasmon-polariton nonlinear structures that further boost the nonlinear refractive index modulation in 30\,nm-thick ITO nanolayers coupled to silicon dioxide/silicon nitride (SiO$_{2}$/SiN) multilayer structures fabricated by magnetron sputtering (MSP). In particular, we 
utilize variable-angle spectroscopic ellipsometry (VASE) to accurately retrieve the
optical dispersion properties of the deposited materials over a large spectral range and we perform Z-scan measurements as a function of the pump intensity in order to precisely quantify their Kerr optical nonlinear coefficients. Our findings unambiguously demonstrate non-perturbative refractive index variations as large as $\Delta{n}\approx{2}$ driven by the near-field coupling of ENZ nanolayers with strongly confined plasmon-polariton Tamm states, providing a stepping stone for the engineering of more efficient nonlinear devices and nanostructures with applications to all-optical data processing, spectroscopy, sensing, and novel infrared photodetection modalities.

\section{Results and Discussion}
\subsection{Field-corrected Susceptibility and Non-perturbative Index Change}

The objective of this section is to concisely review in a pedagogical fashion the perturbative treatment of the induced nonlinear polarization of a generic Kerr medium with intensity dependent refractive index and introduce the necessary modifications needed to describe low-index ENZ materials. We assume for simplicity a centrosymmetric material (i.e., scalar susceptibility) and neglect its magnetic response. Under these assumptions, the total polarization induced by monochromatic light can be written as: 
\begin{equation}
P^{tot}(\omega)=P^{(1)}(\omega)+P^{(3)}(\omega)+\ldots=\epsilon_{0}\left[\chi^{(1)}(\omega)+3\chi^{(3)}(\omega)|E(\omega)|^{2}+\ldots\right]E(\omega)
\end{equation}
where $E(\omega)$ is the incident electric field, $\chi^{(1)}$ is the linear susceptibility of the medium, and $\chi^{(3)}$ is its lowest-order nonlinear susceptibility. From the above expression 
we introduce the ``field-corrected" susceptibility \cite{butcher1990elements}:
\begin{equation}
	\chi_{fc}(\omega)\equiv
	\chi^{(1)}(\omega)+3\chi^{(3)}(\omega)|E(\omega)|^{2}+\ldots
\end{equation}
and the ``field-corrected", or the effective dielectric function:
\begin{equation}
	\epsilon_{fc}(\omega)=1+\chi_{fc}(\omega)
\end{equation}
The linear refractive index is $n_{0}(\omega)=\sqrt{1+\chi^{(1)}(\omega)}
=\sqrt{\epsilon_{r}^{(1)}(\omega)}$ and $\epsilon_{r}^{(1)}$ is the linear contribution to the relative permittivity. In conventional nonlinear materials, susceptibility terms beyond the third-order are negligible in magnitude and the light induced refractive index change $\Delta{n}=n-n_{0}$ is very small under the usual assumption $\Delta{n}<<n_{0}$. Therefore, in these cases the change in  refractive index induced by the incident field can be approximated as follows:
\begin{equation}
	\Delta{n}(\omega)=\sqrt{\epsilon_{fc}(\omega)}
	-n_{0}(\omega)\approx\frac{3\chi^{(3)}|E|^{2}}{2n_{0}(\omega)}
\end{equation}
where we made use of the binomial expansion. The approximate expression above leads to the conventional definition of the intensity dependent refractive index:
\begin{equation}\label{nlindex}
n=n_{0}+n_{2}I
\end{equation}
where:
\begin{equation}
	n_{2}=\frac{3\chi^{(3)}}{4\epsilon_{0}{c}n^{2}_{0}}
\end{equation}
is the nonlinear refractive index of the medium.

It has been recently established by Reshef et al. \cite{Reshef2017}  that when dealing with low-refractive index ENZ media the conventional expression  for the intensity-dependent refractive index in Equation (\ref{nlindex}) is inapplicable as the nonlinear field dependence of the refractive index cannot be expressed in a perturbation series. Instead, the full non-perturbative expression below must be considered in the analysis of ENZ nonlinear media \cite{Reshef2017}:
\begin{equation}\label{nonpert}
	{\tilde{n}(\omega)}=\sqrt{\epsilon_{fc}(\omega)}=\sqrt{\epsilon_{r}^{(1)}+\sum_{j({odd})\neq{1}}c_{j}\chi^{(j)}|E(\omega)|^{j-1}}
\end{equation} 
where we introduced the complex nonlinear refractive index $\tilde{n}(\omega)\equiv{n+i\kappa}$ to emphasize that the quantities in Eq. (\ref{nonpert}) are generally complex and $c_{j}$ are the degeneracy factors associated to each relevant nonlinear process considered in the summation. Specifically, for ITO materials excited at large pump intensity up to $\approx$ 250 GW/cm$^{2}$, it was shown that nonlinear susceptibility up to seventh-order should be included in Equation \ref{nonpert} as their contributions exceed the linear refraction term associated to $\chi^{(1)}$ \cite{Reshef2017}.

Our goal is now to introduce the general analytical expressions for the real and imaginary parts of the complex nonlinear refractive index $n(\omega)$ that will be used in the analysis of the experimental data. In order to achieve this, we first recall the definitions that link the real and imaginary parts of the dielectric function $\epsilon$ of any material with the real ($n$) and the imaginary parts ($\kappa$) of  the complex refractive index $\tilde{n}$, which are shown below:
\begin{subequations}
	\begin{eqnarray}\label{BH}
		n &=& \sqrt{\frac{|\epsilon|+\epsilon_{1}}{2}}\\
		\kappa &=& \sqrt{\frac{|\epsilon|-\epsilon_{1}}{2}} \label{BHb}\\
		|\epsilon|
		&=&\sqrt{\epsilon_{1}^{2}+\epsilon_{2}^{2}}\label{BHc}
	\end{eqnarray}
\end{subequations}
Here $(\epsilon_{1},\epsilon_{2})$ we denote the real and imaginary parts of the field-corrected dielectric function $\epsilon_{fc}$ of an arbitrary nonlinear material and $(n,\kappa)$ are its field-corrected real refractive index and extinction coefficient describing the refractive and the absorptive nonlinearities, respectively. We also recall that $\kappa$ is directly related to the absorption coefficient according to $\alpha=2k_{0}\kappa$, where $k_{0}=2\pi/\lambda$ is the free-space wave number.
The real and imaginary parts of the complex refractive index are obtained by inserting the appropriate perturbative terms of the field-corrected dielectric function in the generally valid expressions listed above. In particular, in our experimental work we focus on low excitation intensities up to $\approx$ 1 GW/cm$^{2}$ where only third-order susceptibility terms are important. We then obtain the following formulas that account for the nonlinear refractive index and the nonlinear absorption coefficient:
\begin{subequations}\label{BH2}
\begin{eqnarray}
n &=& \sqrt{\frac{|\epsilon|+\epsilon^{(1)}_{1}+3\chi^{(3)}_{1}|E|^{2}}{2}}\label{BH2_n}\\
\alpha &=& \frac{4\pi}{\lambda}\sqrt{\frac{|\epsilon|-\epsilon^{(1)}_{1}-3\chi^{(3)}_{1}|E|^{2}}{2}}\label{BH2_kappa}
\end{eqnarray}
\end{subequations}
where: 
 \begin{equation}
 |\epsilon| = \sqrt{(\epsilon^{(1)}_{1}+3\chi^{(3)}_{1}|E(\omega)|^{2})^{2}+(\epsilon^{(1)}_{2}+3\chi^{(3)}_{2}|E(\omega)|^{2})^{2}}
 \end{equation}

\begin{figure}
  \includegraphics[width=\linewidth]{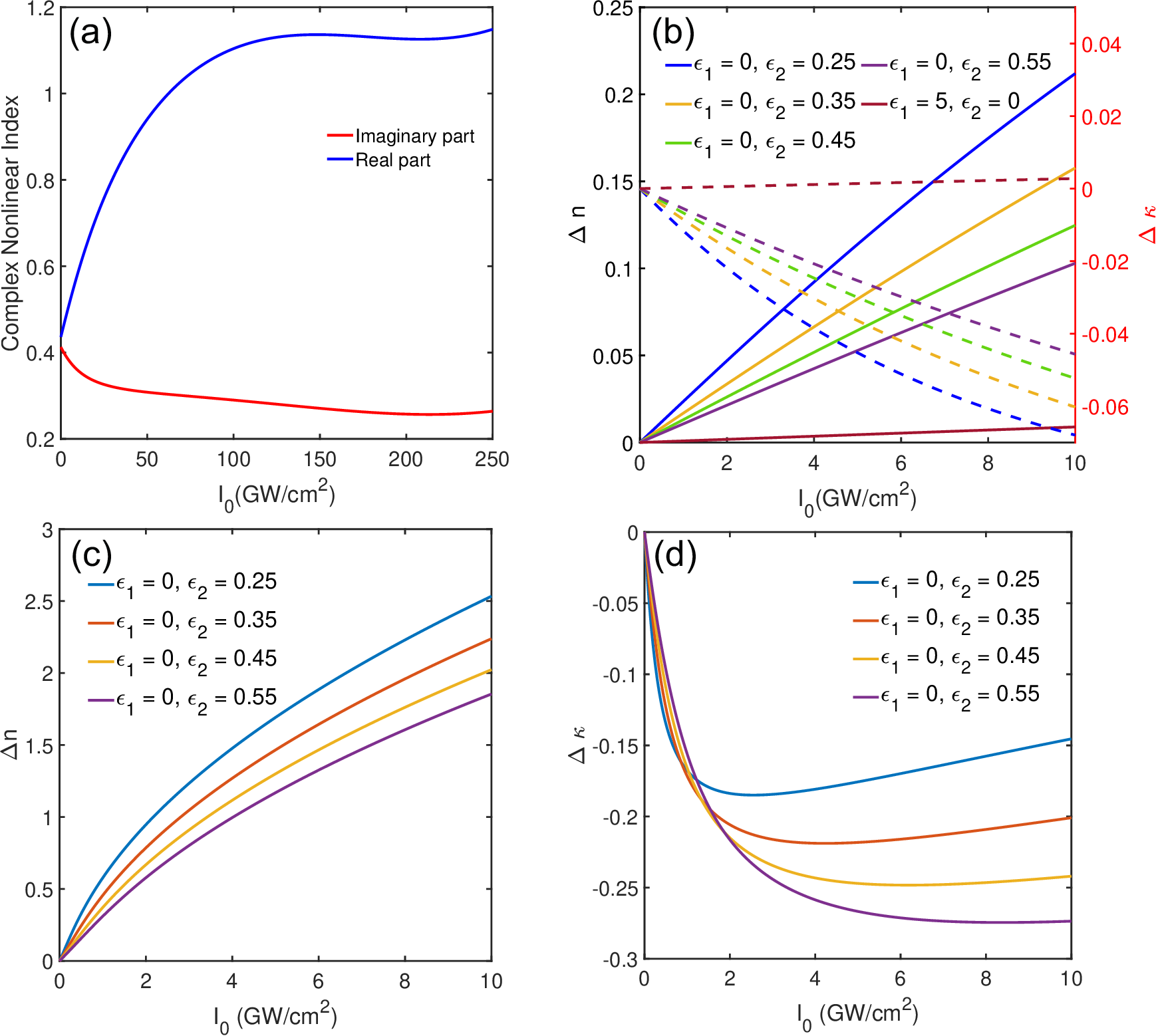}
  \caption{Simulated pump-intensity dependence of the nonlinear refractive index and extinction coefficient of ENZ materials. (a) The complex nonlinear index of ITO as a function of pump power with the susceptibility values of ${\chi}^{(1)}(\frac{m}{V})$=$-$0.98+$i$0.36, ${\chi^{(3)}}$ $(\frac{m^2}{V^2})$=1.6$\times10^{-18}$+ $i$0.5$\times10^{-18}$, ${\chi^{(5)}}$$(\frac{m^4}{V^4})$=-0.63$\times 10^{-36}$ - $i$0.25$\times10^{-36}$ and ${\chi^{(7)}}$$(\frac{m^6}{V^6})$=7.7$\times 10^{-56}$ + $i$3.5$\times10^{-56}$ as measured in ref. \cite{Reshef2017}. (b) Pump-intensity dependence of the nonlinear variations of the refractive index and extinction coefficient for different values of the imaginary components of the linear permittivity $\epsilon_2$ with the same higher-order susceptibilities as in panel (a). (c-d) Nonlinear refractive index and extinction changes, respectively, as a function of pump intensity for the linear permittivity values specified in the legend and considering $\chi^{(3)}(\frac{m^2}{V^2})$=$5.2 \times 10^{-17} + i6\times10^{-18}$ as measured in Ref. \cite{Britton2022}.}
  \label{fig:nonperturbative_example}
\end{figure}

Here  $(\chi_{1},\chi_{2})$ denote, respectively, the real and imaginary parts of the susceptibility function.
We should note that the term $|\epsilon|$ appearing in the above expressions shows how the imaginary parts $\chi^{(3)}_{2}$ and $\epsilon^{(1)}_{2}$ contribute to $n$ and thus give rise to nonlinear refractive effects. Similarly, we see that the real parts $\chi^{(3)}_{1}$ and $\epsilon^{(1)}_{1}$ generally contribute to the nonlinear absorption as well. Interestingly, the manifest interplay between real and imaginary components of the linear/nonlinear parameters of the medium give rise to nontrivial saturation effects as a function of the excitation pump power. We used the generally valid analytical formulas introduced above to model the experimentally measured intensity-dependent nonlinear refractive index and extinction data obtained from the Z-scan technique (see sections \ref{nonlinear1} and \ref{nonlinear2}). 
It is clear that this predictive framework can be naturally generalized to any desired order in the perturbation expansion of the susceptibility. This becomes particularly important when studying ENZ materials at large pump intensities, as illustrated in Figure \ref{fig:nonperturbative_example} (a), where we show the simulated complex nonlinear refractive index of ITO with susceptibility contributions included up to the seventh order. These results are in excellent agreement with the intensity-dependent experimental data reported in reference \cite{Reshef2017}, where the measured values of the nonlinear susceptibility terms of ITO are utilized.

In order to clearly illustrate the benefits of low-index ENZ materials for nonlinear optics applications, we apply our framework to compare in Figures \ref{fig:nonperturbative_example} (b-d) the nonlinear response of Kerr media with optical parameters that are typical of high-quality ITO films at their ENZ transition wavelengths with respect to traditional nonlinear dielectrics. In particular, Figure \ref{fig:nonperturbative_example} (b)  demonstrates the importance of low refractive index media for enhancing nonlinear index and absorption changes. Here, we computed the nonlinear index change $\Delta{n}=n(I_{0})-n_{0}$ (solid lines) where $I_{0}$ is the incident intensity. The field amplitude inside the medium is calculated from the standard formula $I_{0}=2\text{Re}\{n^{(1)}\}\epsilon_{0}c|E|^{2}$ where $n^{(1)}$ is the linear complex refractive index and here $n_{0}\equiv{n_{1}^{(1)}}$. The corresponding nonlinear extinction  coefficient $\Delta{\kappa}=\kappa(I_{0})-\kappa_{0}$ is displayed by the corresponding dashed lines where the largest absorption saturation effects can be clearly observed only in the smallest refractive index regime. Note in contrast the almost complete absence of nonlinear absorption changes for the dielectric medium with $\epsilon_{1}=5$ in the investigated low excitation regime.
Finally, in Figures \ref{fig:nonperturbative_example} (c-d) we show the computed intensity-dependent nonlinear refractive index and extinction changes corresponding to the measured nonlinear susceptibility parameters of the ITO films from reference \cite{Britton2022}. Here, we consider different values of the imaginary part of the linear permittivity and observe in panel (c) how both the slope and the total refractive index change are maximized by the materials with the lowest values of $\epsilon_{2}$. Interestingly, panel (d) also shows the remarkable effect of $\epsilon_{2}$ on the nonlinear extinction coefficient, where materials with the lowest value of $\epsilon_{2}$ display a characteristic``saturation reversal" of $\Delta\kappa$ leading to a gradual recovery of the linear extinction at the largest pump intensities.
We found the interplay of $\epsilon_2$ with the linear losses in the low-index regime leads to a characteristic inversion regime in which the materials switch from loss saturation to enhanced absorption by increasing the pump intensity. Therefore, both the saturation and enhancement of the extinction coefficient can be observed as a function of pump intensity in low-index ITO materials characterized by the positive $\chi^{(3)}$ values previously reported in literature \cite{Reshef2017,Britton2022}.

\subsection{Linear Optical Dispersion of ITO Thin Films}
\begin{figure}
  \includegraphics[width=\linewidth]{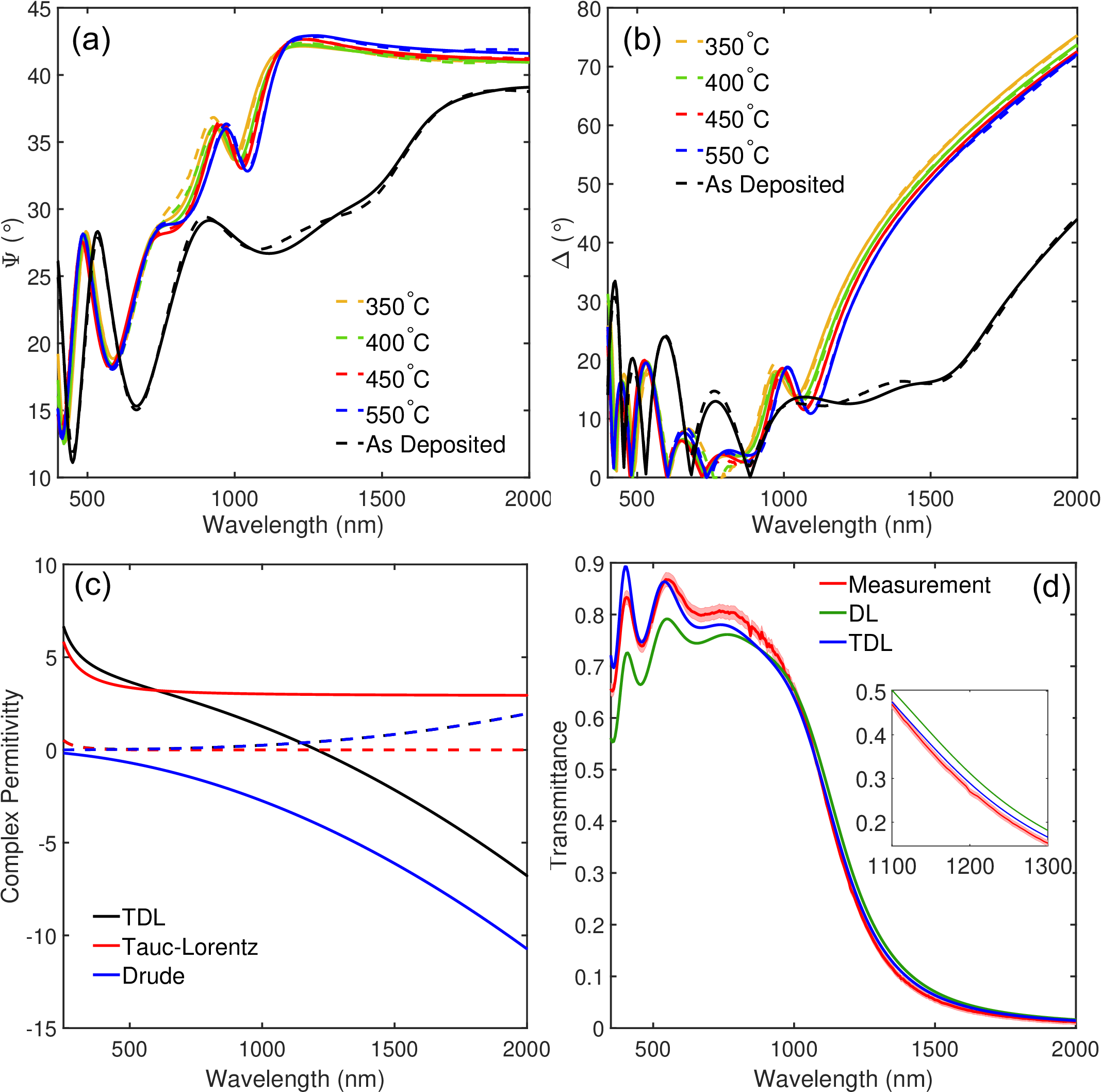}
  \caption{Characterization of the linear optical properties of fabricated ITO films. Representative (a) $\Psi$ and (b) $\Delta$ ellipsometry data (dashed) and fit (solid) of 300\,nm ITO thin films for various post-deposition vacuum annealing temperatures. (c) Representative complex permittivity of ITO annealed at 400$^\circ$C separated into the components of the Drude and TL oscillators used to fit the experimental data along with the combined TDL complex permittivity used to fit the data over a broad range. The solid lines and dashed lines represent the real and imaginary components of each oscillator respectively. (d) Measured normal-indicence transmission spectrum of the ITO sample annealed at 400$^\circ$C and best-fits  using the DL and TDL ellipsometry models. Error bars are visible in the shaded red region.}
  \label{fig:linear_ellipsometry}
\end{figure}

In this section we discuss the linear optical properties of low-index ITO films. In particular, we use variable angle spectroscopic ellipsometry (VASE), which is an accurate technique largely utilized for the characterization of the linear dispersion properties of materials \cite{aspnes2014spectroscopic}. This enables the development of accurate dispersion models over a broad wavelength range, which is key for the unambiguous determination of the linear optical constants of materials. In spectroscopic ellipsometry, a parameterized oscillator model is introduced to capture the relevant aspects of optical dispersion as a function of wavelength, to interpret the measured data, and to extract the optical constants of materials via accurate fitting procedures \cite{woollam_ellips1,woollam_ellips2,Johs2008}. The accuracy of the obtained optical constants is then tested through fitting of independent transmission/reflection data.  Here, we address the linear optical response of ITO thin films over a wide spectral range extending from the ultraviolet (UV) to the near infrared (NIR) wavelengths. We utilized broadband VASE and normal incidence transmission measurements to accurately retrieve the optical dispersion properties of 300\,nm-thick films of ITO grown atop of fused silica substrates. The details of the ITO growth by radio-frequency magnetron sputtering can be found in section 4. VASE measurements at three angles (65$^\circ$, 70$^\circ$, 75$^\circ$) were performed directly on the 
double-side polished fused silica substrate using the back-surface reflection suppression technique \cite{backsurfacereflection}. Figure \ref{fig:linear_ellipsometry}(a) and \ref{fig:linear_ellipsometry}(b) show the measured and fitted ellipsometric parameters $\Psi$ and $\Delta$ corresponding to the magnitude and phase of the ratio of the complex s- and p-polarized Fresnel reflection coefficients, respectively, at an angle of $75^\circ$, for annealed and as-deposited samples. Ellipsometry measurements performed at 65$^\circ$ and 70$^\circ$ along with fits of each fabricated ITO sample are shown in the supplemental material. The complex Fresnel reflection coefficients are then related to the complex permittivity and thickness of the material through parameterized oscillator models\cite{fujiwara2007spectroscopic}. Due to the conductivity of ITO and its transparency in the UV and visible (VIS) wavelengths, the traditional approach to dispersion modeling of ITO thin films is based on a combination of Drude and Lorentz oscillators\cite{2020APL_RodriguezSune, Scalora2020}. However, recent work has established the need to extend the Drude-Lorentz (DL) dispersion model in order to accurately describe the complex permittivity of ITO across a broad spectral range \cite{Britton2022, Gui2019,Synowicki1998_ITO}. In fact, while the DL model well accounts for a direct optical bandgap (E$_g$), it also overestimates the optical losses due to the ``long-tailed" nature of the Lorentzian functions involved \cite{Synowicki1998_ITO}. To avoid this problem, here we employ the Tauc-Drude-Lorentz (TDL) oscillator model that is consistent with Kramers-Kronig causality relations \cite{Britton2022, Gui2019,Synowicki1998_ITO}. 
In particular, the Tauc-Lorentz (TL) oscillator model, which combines the Tauc joint density of states with the standard quantum mechanical Lorentz oscillator, has been introduced to parametrize the optical constants of amorphous materials in excellent agreement with experimental measurements \cite{jellison2000characterization,Jellison1996}.

The permittivity function of the TDL model used in our paper is as follows:
\begin{equation}
{\epsilon}^{TDL} = 1 + \epsilon^{Drude} + \epsilon_{1}^{TL} + i\epsilon_{2}^{TL}
\end{equation}
where:
\begin{subequations}
	\begin{align}\label{Drude TL Model}
        \epsilon^{Drude}(E) &=\frac{-\hbar}{\epsilon_0\rho(\tau*E^2 + i\hbar E)}&\\
        \epsilon_{2}^{TL}(E) &=\frac{AE_{0}C(E - E_g)^2}{(E^2-E_{0}^2)^2 + C^2E^2}\cdot\frac{1}{E} \quad &\mathrm{for} \, E > E_{g}\\
        \epsilon_{2}^{TL}(E) &= 0  &\mathrm{for} \, E < E_g\\
        \epsilon_{1}^{TL}(E) &= \frac{2}{\pi}P\int_{E_g}^{\infty}\frac{\xi\epsilon_{2}^{TL}(\xi)}{\xi^2 - E^2}d\xi \label{TL_Int}
	\end{align}
\end{subequations}
Here $\rho$ and $\tau$ are free parameters of the Drude model corresponding to the resistivity and scattering time, respectively, and $A$, $E_{0}$, $C$, $E_g$ are free parameters of the TL oscillator corresponding to the amplitude, center energy, broadening, and the band gap energy respectively. The symbol $P$ denotes the Cauchy principal value integral in equation 11d. This integral can be evaluated exactly, leading to a long closed-form expression that can be found in references \cite{Jellison1996,Jellison1996_erratum}. The different contributions to the complex permittivity of the Drude model, TL, as well as the TDL model used here are separately displayed in figure \ref{fig:linear_ellipsometry}(c). In table \ref{table:ITO_model_params} we report the oscillator parameters of each fabricated ITO sample used to fit the ellipsometric measurements as well as the mean square error(MSE) values for each fit\cite{woollam_ellips1}.

\begin{table}
    \centering
    \begin{tabular}{ || c c c c c c c c c || }
    \hline
    T ($^\circ$C) & $\rho(\Omega\cdot cm)$ & $\tau(fs)$ & $A(eV)$ & $E(eV)$ & $C(eV)$ & $E_g(eV)$ & MSE & $\lambda_{ENZ}(nm)$ \\ [0.5 ex]
    \hline \hline
    $350$ & $1.9\times 10^{-4}$ & 5.76 & 44.95 & 6.99 & 0.87 & 2.27 & 11.47 & 1150\\
    $400$ & $2.19\times 10^{-4}$ & 5.38 & 45.16 & 6.41 & 0.86 & 2.4 & 10.55 & 1200\\
    $450$ & $2.1\times 10^{-4}$ & 5.78 & 45.16 & 7.17 & 0.91 & 2.12 & 9.88 & 1215\\
    $550$ & $1.8\times 10^{-4}$ & 6.76 & 50.1 & 7.03 & 0.78 & 2.48 & 15.08 & 1235\\
    As Dep. & $6.2\times 10^{-4}$ & 4.1 & 129.8 & 9.122 & 20.6 & 2.97 & 27.36 & 1800\\
    \hline
    \end{tabular}
    \caption{Drude and TL oscillator parameters of each fabricated ITO sample annealed at temperture, $T$, used to fit ellipsometry and transmission measurements along with their respective MSE as defined in reference \cite{woollam_ellips1} and the corresponding ENZ wavelengths of each sample}
    \label {table:ITO_model_params}
    \end{table}

The impact of these different approaches in modeling transmission data is clearly shown in figure \ref{fig:linear_ellipsometry}(d), where we plot the transmission spectrum of a representative ITO sample measured at normal incidence and the best-fits obtained using the DL and the TDL models. We find that the TDL oscillator model results in a better agreement with the data particularly in the UV-VIS region, where a discrepancy of up to 15$\%$ exists using the simpler DL model. Moreover, as we show in the inset of figure \ref{fig:linear_ellipsometry}(d), the DL model yields a sizeable disagreement with measured data also across the NIR region, where ENZ conditions occur in ITO films. Particularly around the ENZ region of interest, inaccurate characterization of the linear complex permittivity of ITO leads to significant inaccuracies when characterizing higher-order nonlinear optical properties. On the other hand, we found that the developed TDL model matches very well the experimental results over the entire spectral region investigated. It should be noted that the fits for the the DL and TDL models are obtained when considering ellipsometric reflection measurements at all three angles as well as the normal incidence transmission measurements simultaneously. This ensures robust identification of the parameters that best reproduce the experimental data over a broad spectral range based on independently measured data sets.

\begin{figure}
  \includegraphics[width=\linewidth]{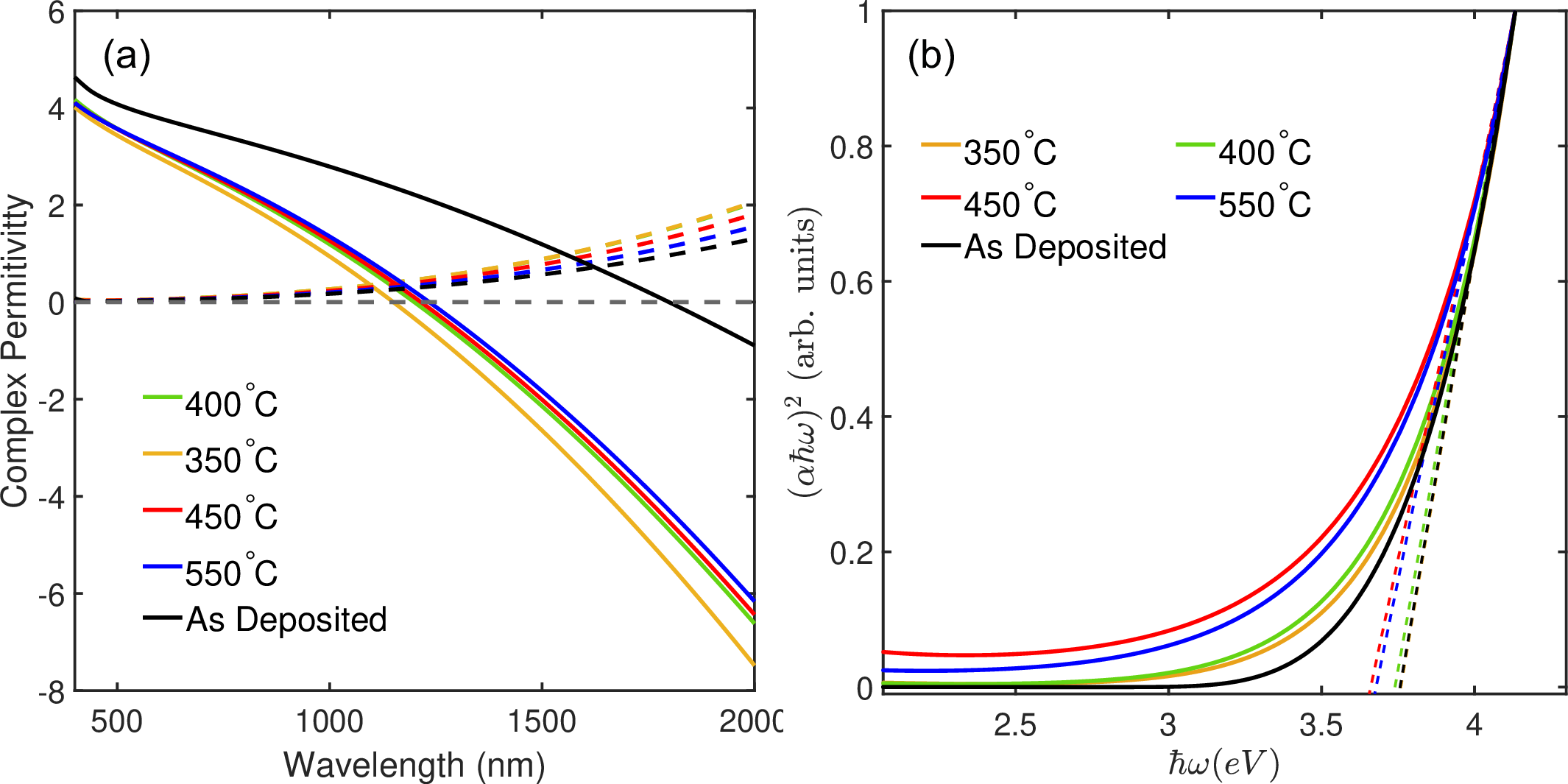}
  \caption{\ (a) Real (solid) and imaginary (dashed) optical dispersion data of the fabricated ITO thin films obtained from the TDL ellipsometric parameters reported in the main text. (b) Tauc plot of the ITO films showing the direct optical band-gap extrapolated from the linear regions (dashed lines).}
  \label{fig:linear_epsilon_tauc}
\end{figure}

We remark that ITO films are commonly grown with a graded micro-structure that lead to variations of the optical constants as a function of film thickness \cite{Gui2019}. We account for this variation through a linear grading of E$_g$ and the Drude model parameters $\rho$, $\tau$ as a function of film thickness. We find that this linear grading method, which plays a small role in the overall change of permittivity with variations of $\Delta \tilde{\epsilon} <0.1$ throughout the thickness of the ITO, still improves the quality of the fits to the experimental data decreasing the MSE from a value $\approx 25$ without any grading to a value $\approx10$. The extracted complex permittivity of each fabricated sample is plotted in Figure \ref{fig:linear_epsilon_tauc}(a) and the Tauc plots that identify the direct band-gaps from the measured optical constants are shown in Fig. \ref{fig:linear_epsilon_tauc}(b). From these data we find that post-deposition annealing strongly blue-shifts the ENZ region to $\lambda\approx 1200$\,nm due to the improved defect passivation upon  the crystallization of ITO thin films, as previously shown in \cite{Britton2022}, while simultaneously reducing losses and increasing conductivity with respect to the as-deposited sample. We also remark that the direct optical band-gap of each fabricated sample is $\approx 3.7$eV, consistently with published results \cite{Wang2015,Hamber_ITO_bandgap} . In the following section, we turn our attention to the characterization of the nonlinear optical properties of the fabricated ITO materials in the ENZ regime.

\subsection{Nonlinear Refractive Index Change in the Epsilon-Near-Zero regime} \label{nonlinear1}
\begin{figure}
  \includegraphics[width=150mm]{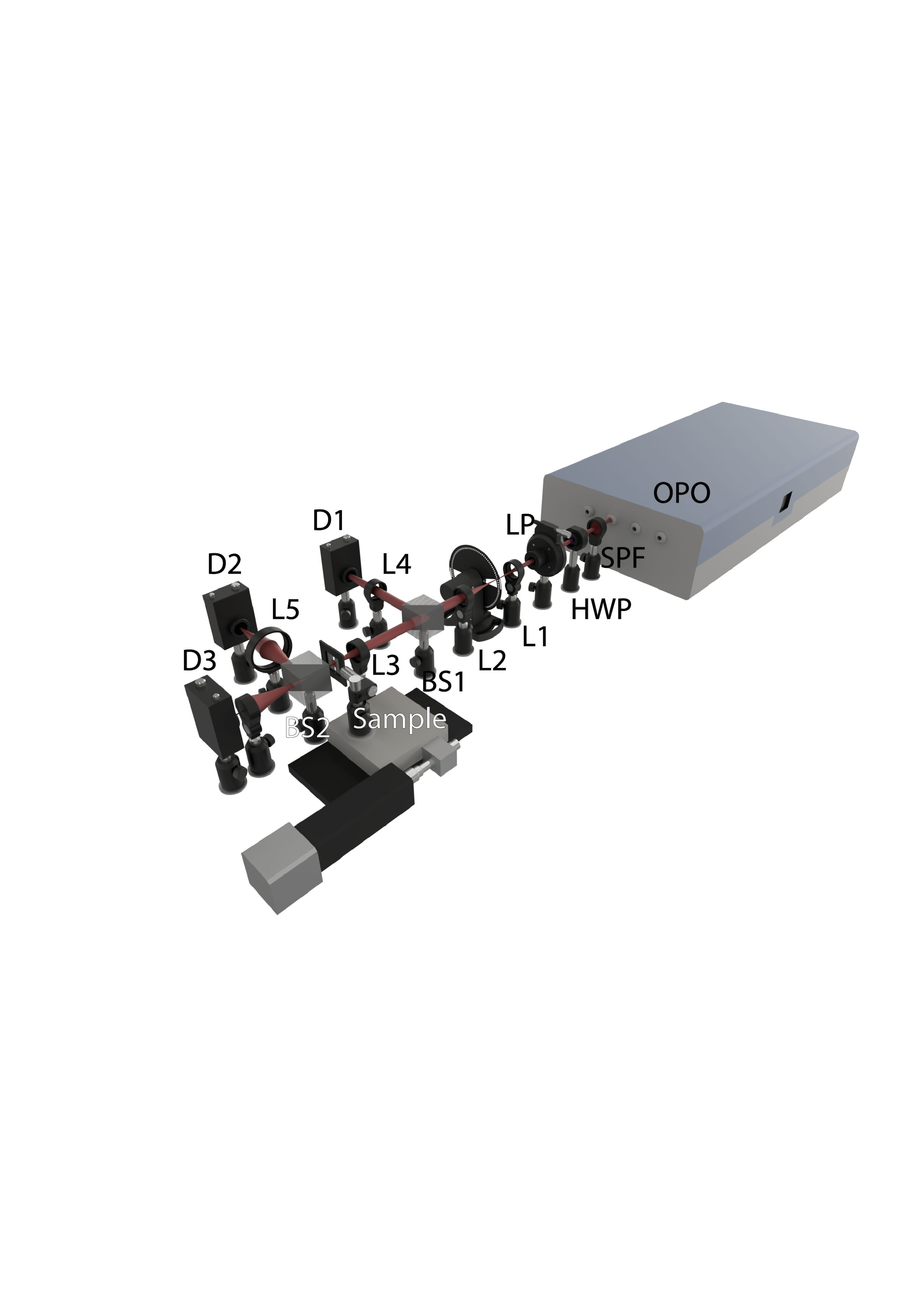}
  \caption{Z-scan experimental setup enabling simultaneous closed- and open-aperture measurements.}
  \label{fig:z-scan setup}
\end{figure}

We characterize the third-order optical nonlinearity of the fabricated ITO thin films via the Z-scan technique \cite{Z-scanOG}. This technique enables the accurate determination of the nonlinear susceptibility of the samples, which in general is a complex quantity, through the precise characterization of the complex refractive index change $\Delta\tilde{n}$ as a function of the pump intensity \cite{Kuzyk1998}. In Figure \ref{fig:z-scan setup} we illustrate the experimental Z-scan setup used in this work to enable simultaneous open- and closed-aperture measurements. Additional details of the setup can be found in section 4. In order to reduce undesired thermo-optical effects, a mechanical chopper with a $10\%$ duty cycle and operation frequency of 150 Hz was utilized, resulting in a 250 $\mu$s rise time. We find no evidence of time-dependent thermo-optical contributions within the sampling time window of $\approx$ 650 $\mu$s of the experimental setup, consistently with our previous studies \cite{Britton2022}. Furthermore, we calculate the expected thermal-lensing contributions to closed-aperture Z-scan measurements under our experimental conditions using the expressions derived in ref. \cite{Shehata2019}. Additional details on this analysis are reported in the supplemental materials.

In Figure \ref{fig:ITO z-scan_results} we summarize the results on the nonlinear optical response of the fabricated ITO thin films with a thickness of 300\,nm. In particular, panels \ref{fig:ITO z-scan_results}(a) and \ref{fig:ITO z-scan_results}(b) show representative open- and closed-aperture Z-scan data, respectively, of ITO annealed at 400$^\circ$C. These measurements are performed as a function of the incident pump intensity and at the fixed ENZ pump wavelength $\lambda = 1200$\,nm. The continuous lines are the theoretical results for both open- and closed-aperture scans obtained by simulating the electric field  propagation in a complex $\chi^{(3)}$ 
nonlinear interface using the Rayleigh–Sommerfeld first integral formulation \cite{goodman2005introduction,Britton2022,Britton_axilens}. The relative intensities of the unobstructed beam or of the beam transmitted by a 200 $\mu$m aperture in the far field are computed for open- and closed-aperture scans, respectively. The electric field response of a complex nonlinear $\chi^{(3)}$ thin layer can be expressed as \cite{Z-scanOG}:
\begin{equation}
    E(z,r) = E_0(z,r)e^{\frac{-\alpha L_{eff}}{2}}\left[1 + \beta I(z,r)L_{eff}\right]\exp{\left(\frac{ikn(I_0)}{\beta} - \frac{1}{2}\right)}
\end{equation}
Here $E_0$ is the initial field at the plane of the interface, $\alpha$ the linear absorption coefficient, $\beta$ the nonlinear absorption coefficient, $n(I_0)$ the nonlinear refractive index, as defined in section 2.1, and $k$ the wave number. $L_{eff}$ is the effective sample thickness defined as $L_{eff} = (1-e^{\alpha L})/\alpha$ where L is the thickness of the sample. All fits are performed using a nonlinear least squares fit solver and the fitting parameter covariance is estimated from the output Jacobian and residuals with the 95\% confidence interval reported. The open-aperture scans, shown in figure \ref{fig:ITO z-scan_results}(a), demonstrate an increase in transmission as the sample is translated through the focused Gaussian beam indicating a reduction in optical losses due to saturable absorption, consistently with the reports in reference \cite{Alam2016,Britton2022}. Notice that this effect is also evidenced in the closed-aperture scans, shown in figure \ref{fig:ITO z-scan_results}(b), from the asymmetry between the peak and trough that grows proportionally to the peak observed in the open-aperture scans\cite{Z-scanOG}. The high quality of the fits for both open- and closed-aperture scans allows us to unambiguously estimate the change in complex refractive index as a function of peak pump power through the  relations $\Delta \kappa = \beta I_0/2k$ and $\Delta n = \Delta \Phi/(kL_{eff})$ where $\Delta \Phi$ is the measured nonlinear phase shift at the focus of the beam. We adjust $I_0$ as necessary to account for the $n_0$-dependent Fresnel reﬂection losses at the ﬁlm/air interface by computing the transmission/reflection properties of the thin film. These parameters enable the determination of the intrinsic third-order nonlinear susceptibility using the generally valid equations \ref{BH2}(a-c). The extracted change in extinction and refractive index as a function of pump power along with the corresponding fits for each annealed ITO sample are reported in figure \ref{fig:ITO z-scan_results}(c) and \ref{fig:ITO z-scan_results}(d) respectively. From these results, we observe a significant enhancement in the optical nonlinearity of the ITO sample annealed at $400^\circ$C with non-perturbative saturation effects in both the real and imaginary components of the complex nonlinear refractive index trends in excellent agreement with their simulated behavior obtained from the equations \ref{BH2_n} and \ref{BH2_kappa} (solid lines). Furthermore, we find that at powers of $\approx$ 1 GW/cm$^{2}$, the nonlinear refractive index change $\Delta n$ exceeds the linear index $n_0$ for all but one of the samples investigated. In fact, the ITO sample annealed at the highest temperature ($550^\circ$C) showed a substantially lower refractive index change with no measureable open-aperture signal, consistently with our previous findings \cite{Britton2022}. The enhanced nonlinear response of the samples annealed at $400^\circ$C is attributed to the development of secondary phases with anisotropic crystallographic texturing in sputtered ITO films \cite{Britton2022}. We summarize in Table \ref{table:ITO_Chis} the values of the measured nonlinear coefficients for each annealed sample, demonstrating order-of-unity refractive index changes in the fabricated materials. Note that the sign of the imaginary component $\chi^{(3)}_2$ is positive for all reported values despite the observed reduction in the optical losses. This feature has also been reported by Reshef et al. \cite{Reshef2017} for ITO films and is a direct manifestation of the interplay between the linear and nonlinear complex susceptibilities of low-index materials previously highlighted in relation to equation \ref{BH2}. To the best of our knowledge, this characteristic behavior is another unique feature of the nonlinear response of low-index ITO films at their ENZ wavelengths. However, despite the measured ($\approx$15\%) nonlinear reduction in optical losses at pump powers of $\approx$ 1 GW/cm$^{2}$, ITO films with 300\,nm thickness are still too lossy to be considered a viable platform for integrated nonlinear applications \cite{Reshef2019_ENZ_Review}. Considerations of ITO for practical nonlinear devices suggest that propagation length and film thickness must be greatly reduced in order to compensate for the intrinsic losses of these materials. Fortunately, it has been shown that ultra-thin (less than $\lambda/50$ in thickness) films of ITO support strongly confined modes close to their ENZ wavelengths \cite{ENZModes_2015}, called Berreman modes \cite{BerremanModes_2015,Vassant:12,Bello2017}, which give rise to a considerable field enhancement. However, these modes cannot be excited from free space and require extrinsic coupling mechanisms. In the next section, we propose a strategy for overcoming this difficulty and significantly enhancing the nonlinear response of ultra-thin (30nm-thick) ITO nanolayers based on the coupling with Tamm polariton surface states in compact nonlinear structures.

\begin{table}
    \centering
    \begin{tabular}{ || c c c c c c || }
    \hline
    T(C) & $n_0$ & $\kappa_0$ & $\chi^{(3)}_1(m^2/V^2)$ & $\chi^{(3)}_2(m^2/V^2)$ & $\Delta n_{max}/n_0$ \\ [0.5 ex]
    \hline \hline
    $350^\circ$ & 0.34 & 0.67 & $4.25\times10^{-17}$ & $7.89\times10^{-18}$ & 1.14\\
    $400^\circ$ & 0.47 & 0.48 & $7.58\times10^{-17}$ & $2.9\times10^{-17}$ & 1.08 \\
    $450^\circ$ & 0.41 & 0.52 & $5.14\times10^{-17}$ & $1.05\times10^{-17}$ & 1.01 \\
    $550^\circ$ & 0.56 & 0.31 & $5.14\times10^{-17}$ & - & 0.44\\
    \hline
    \end{tabular}
    \caption{Third order nonlinear susceptibility coefficients, $\chi^{(3)}$, of each ITO sample annealed at temperature, T, extracted from pump power dependent z scans. The respective linear refractive indices at the pump wavelength $\lambda$=1200\,nm and maximum relative change in refractive index is also shown.}
    \label {table:ITO_Chis}
    \end{table}

\subsection{Enhanced Nonlinear Response of Tamm Plasmon-polariton Structures} \label{nonlinear2}
Surface plasmon-polaritons (SPPs) are hybrid excitations supported by metal-dielectric interfaces that have been utilized with great success to localize and enhance light-matter interactions at the sub-wavelength scale \cite{Bello2017,dielectricnanosphere_plasmon2006}. Conventionally, a SPP is formed with a TM-polarized beam at the boundary of a metallic and dielectric material and features a dispersion curve that lies below the light-line, rendering free space optical excitation impossible \cite{novotny_hecht_2012}. Therefore, diffraction gratings \cite{Maier2005} or bulky dielectric prisms on a thin metal layer \cite{KretschmannRaether_1968} are traditionally used to excite surface waves, which is not always practical for device applications. In order to overcome these limitations, a special type of surface polariton state, known as Tamm plasmon-polariton (TPP), was recently proposed \cite{Kaliteevski2007} and experimentally demonstrated in 2008 by Sasin et al. \cite{Sasin2008}. This surface wave is the optical analog of the electronic Tamm states that form on the surface of a crystalline material due to the breaking of its translational symmetry \cite{Tamm_original}. In recent times, photonic TPP structures have been utilized for the demonstration of tunable perfect absorbers for VIS-NIR wavelengths \cite{Lu:17, Lu:16,Liu:17, Gong:11,nano11123447}, Tamm lasers with tailored emission properties \cite{GaAs/Silver_Tamm_Laser,UVTamm_Laser_2022,Confined_TPP_laser_2013}, thermal emitters \cite{TPP_Thermal_2020,TPP_thermal_2018}, temperature sensors \cite{KUMAR201710,AHMED2021114387,Tamm_temp1} and real time chemical/bio sensors \cite{coatings10121187,Tamm_chem2,Tamm_chem1,LI2018644}. Photonic TPP structures are composed of a distributed Bragg reflector (DBR) terminated by a thin metal film. Unlike traditional SPPs, in these structures TPP surface states are formed within the optical gap of the DBR. TPPs have a number of desirable optical properties that are well-suited for integrated nonlinear nano-optics. Specifically, the dispersion curve of these surface modes is entirely above the free photon light-line enabling broadband free space optical excitation in a compact structure. Moreover, TPP modes can be excited using either TE or TM polarized light at any incident angle, and feature a parabolic band dispersion similar to the one of free massive particles. Finally, it has been shown that TPP modes exhibit ultra-fast response times, on the order of a few femtoseconds, significantly outpacing traditional SPPs \cite{Afinogenov2016}. 

\begin{figure}
  \includegraphics[width=\linewidth]{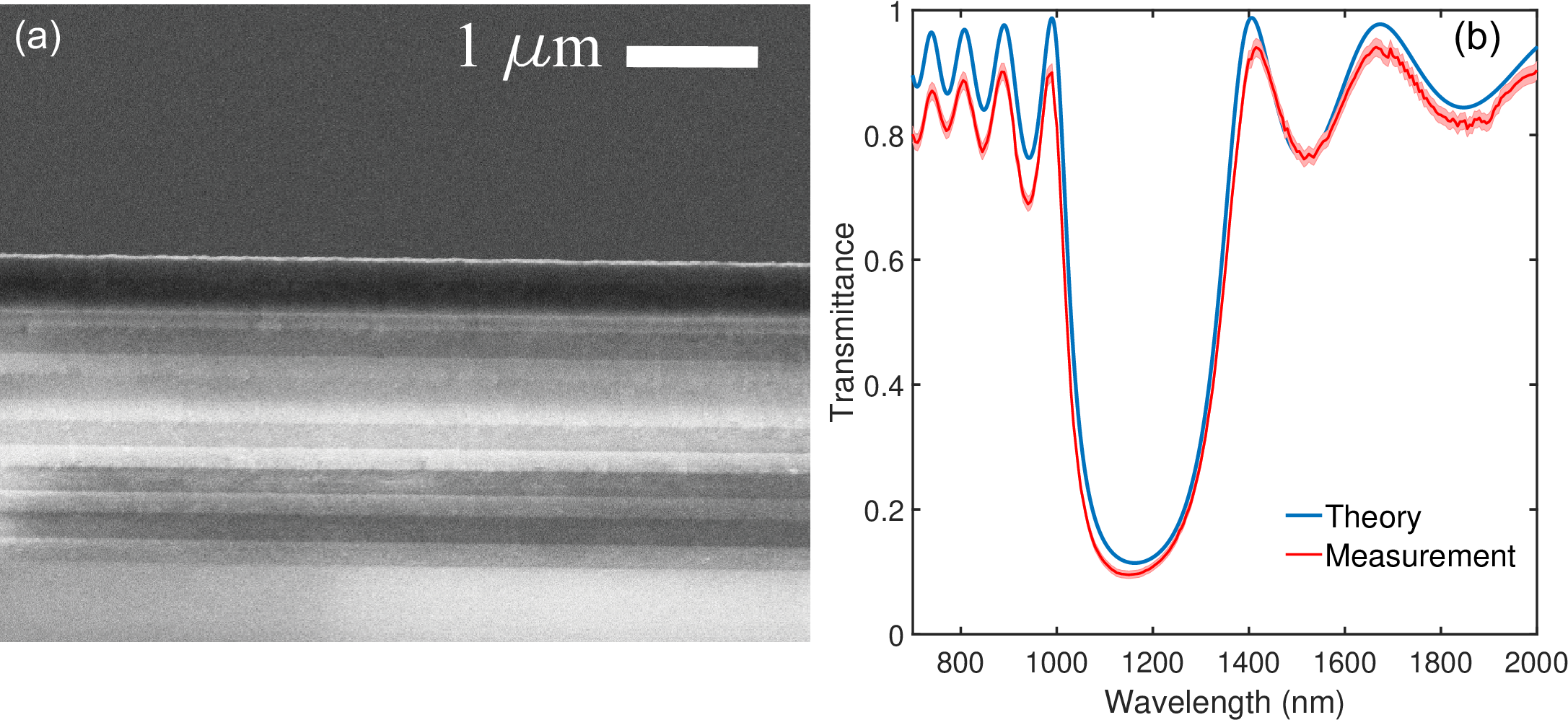}
  \caption{Fabrication and characterization of the distributed Bragg mirror. (a) Cross sectional SEM image of a 13-layer dielectric stack composed of silicon nitride and silicon dioxide grown on top of a quartz substrate. (b) Transmission spectrum of the Bragg mirror in panel (a) measured at normal incidence. Red and blue lines represent measurement and TMM simulations, respectively.}
  \label{fig:bragg}
\end{figure}

In this section we design and fabricate a highly nonlinear TPP structure containing an embedded 30\,nm-thick ITO nanolayer positioned at the location of maximum field intensity enhancement of the TPP surface wave supported by the dielectric DBR structure.  The resonance frequency of the TPP structure can be estimated with the following expression \cite{Kaliteevski2007}:
\begin{equation}\label{eq:DBR_eq}
    \omega_{TP} \approx \frac{\omega_{DBR}}{1+2n_a\omega_{DBR}/n_b\beta\omega_p}
\end{equation}

where $\omega_{DBR}$ is the center frequency of the DBR, $n_a$ and $n_b$ are the refractive indices of the constituent materials, $\beta = {\pi n_a}/{|n_a - n_b|}$ and $\omega_p$ is the plasma frequency of the metal. We note that Equation (\ref{eq:DBR_eq}) requires $n_a > n_b$, where the material with refractive index $n_a$ is adjacent to the metal film. In fact, no TPP mode exists within the first stop band of the DBR if this condition is not met \cite{Kaliteevski2007}. We design the resonance frequency of our TPP  structure to overlap the ENZ wavelength range of the nonlinear ITO nanolayer in order to achieve maximum field enhancement in the spectral region of strong nonlinear response. In order to demonstrate a nonlinear TPP photonic structure, we first fabricate a distributed Bragg reflector (DBR) with 13 layers by growing alternating nanolayers of silicon nitride (SiN) and silicon dioxide (SiO$_2$) atop a fused silica substrate via RF sputtering. The growth parameters of each material are detailed in section 4. Each SiN and SiO$_2$ layer was grown with a nominal thickness of 149\,nm and 198\,nm, respectively, leading to a central stop-band wavelength of $\approx 1190$\,nm. A cross-section SEM image of the fabricated dielectric mirror terminated by a thin layer of gold (for charge dissipation) is shown in figure \ref{fig:bragg}(a). Moreover, a transmission measurement at normal incidence on the fabricated DBR is shown in figure \ref{fig:bragg}(b) along with the simulated transmission using the transfer matrix method (TMM) \cite{wartak2013computational}, indicating excellent agreement with the experimental data within the relevant stop-band region. Finally, a 30\,nm-thick ITO nanolayer was incorporated inside the uppermost SiN layer by a sequential deposition process consisting of the following steps: (i) growth of SiO$_2$ layer with same thickness as previous layers (ii) growth of a SiN layer with half the thickness of the previous layers; (iii) deposition of a 30\,nm ITO nanolayer with the same deposition conditions and post-deposition vacuum annealing as reported earlier, and finally (iv) deposition of the second half of the uppermost SiN layer atop the DBR. Finally, in order to form a TPP state in these nonlinear photonic structure we evaporated a 24\,nm-thin layer of gold via electron beam evaporation using a CHA Industries Solution Process Development System at a base pressure of $1\times10^{-7}$ Torr. 

\begin{figure}
  \includegraphics[width=\linewidth]{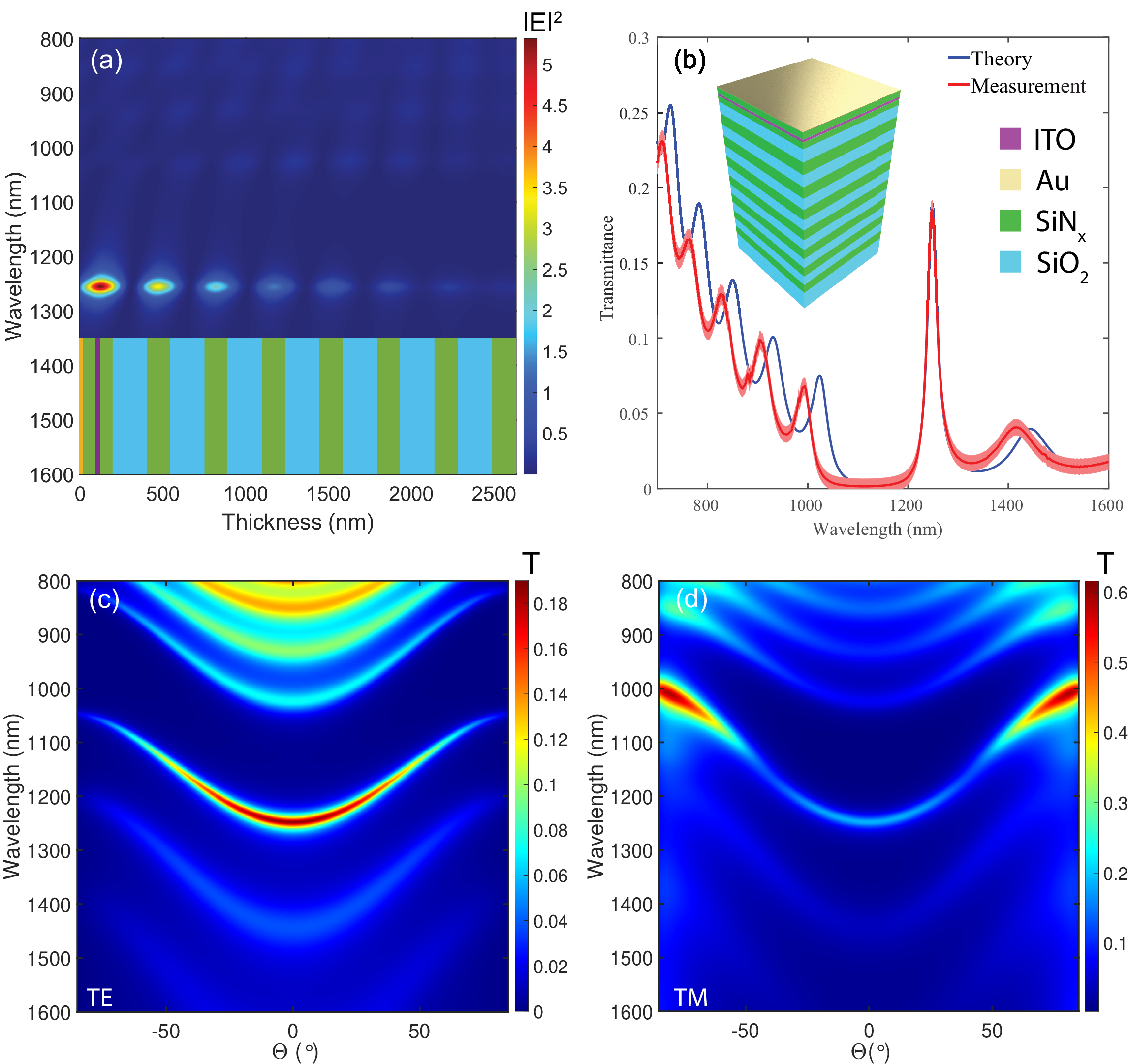}
  \caption{ Design of non-linear Tamm state structure. (a) Field intensity enhancement map of the fabricated nonlinear Tamm state device as a function of wavelength and depth of the multilayer stack. The thickness and material used in each layer of the stack are represented by the colored bars. A 3D schematic of the multi-layer stack is shown in panel (b) along with a color-coded legend for each material used. (b) Transmission measurement of the fabricated device at normal incidence with error bars represented by the red shading and the simulated transmission according to TMM theory (blue line). (c-d) Calculated transmission maps (i.e., transmission as a function of the incidence angle and wavelength) of the multilayer stack for (c) TE and (d) TM polarization.}
  \label{fig:1DTamm_trans}
\end{figure}

\begin{figure}
  \includegraphics[width=\linewidth]{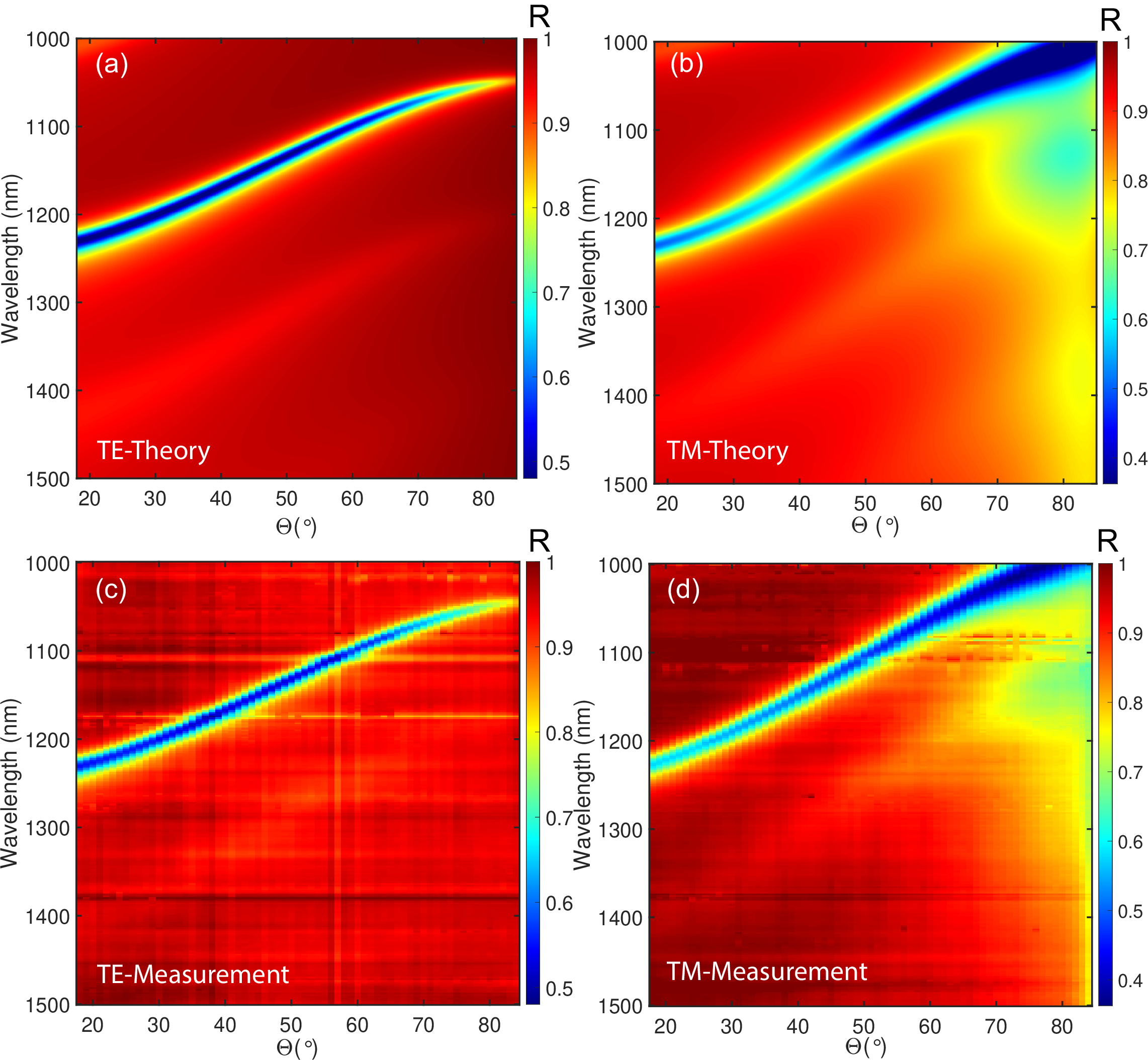}
  \caption{Variable-angle reflection of as a function of wavelength characterizing the dispersion behavior of Tamm states in the fabricated structures. (a-b) Calculated and (c-d) measured reflection maps of the nonlinear Tamm state devices for both TE and TM polarization, respectively.}
  \label{fig:1DTamm_reflec}
\end{figure}

The linear optical properties of the fabricated nonlinear TPP structure were characterized by transmission measurements at normal incidence and variable angle reflection measurements performed with VASE. In figure \ref{fig:1DTamm_trans}(a), we plot the calculated scattering map of the structure \cite{Fibonacci_DalNegro}, which shows the wavelength dependent electric field intensity distribution inside the fabricated structure excited at normal incidence. This analysis demonstrates approximately a 5$\times$ enhancement in the optical intensity spatially overlapping with the 30\,nm ITO layer, as shown by the schematics below. A sketch of the fabricated TPP nonlinear photonic structure is also shown in the inset of figure \ref{fig:1DTamm_trans}(b). In figure \ref{fig:1DTamm_trans}(b), we report the measured and simulated transmission spectra at normal incidence, which demonstrate a resonance peak at 1250\,nm corresponding to the excitation of the TPP surface wave in excellent agreement with the TMM theory. The calculated transmission of the structure as a function of wavelength and incident angle for both TE and TM polarization is shown panels \ref{fig:1DTamm_trans}(c) and \ref{fig:1DTamm_trans}(d), respectively. The data clearly demonstrate a parabolic dispersion characteristic of the surface TPP state. Furthermore, we confirm the polarization dependent dispersion characteristics of our structure through variable angle reflection spectral measurements with an angular resolution of $1^\circ$ and compare them to the theoretical predictions in figure \ref{fig:1DTamm_reflec}. The measurements clearly demonstrate distinct TPP surface modes excited through free space at all the investigated angles and polarizations, in excellent agreement with TMM theory. Moreover, the observed quadratic dispersion behavior of the TPP structures allows us to fine tune the resonance wavelength of the structure and match the ENZ condition of the embedded ITO nanolayer by a small angular adjustment. 

\begin{figure}
  \includegraphics[width=\linewidth]{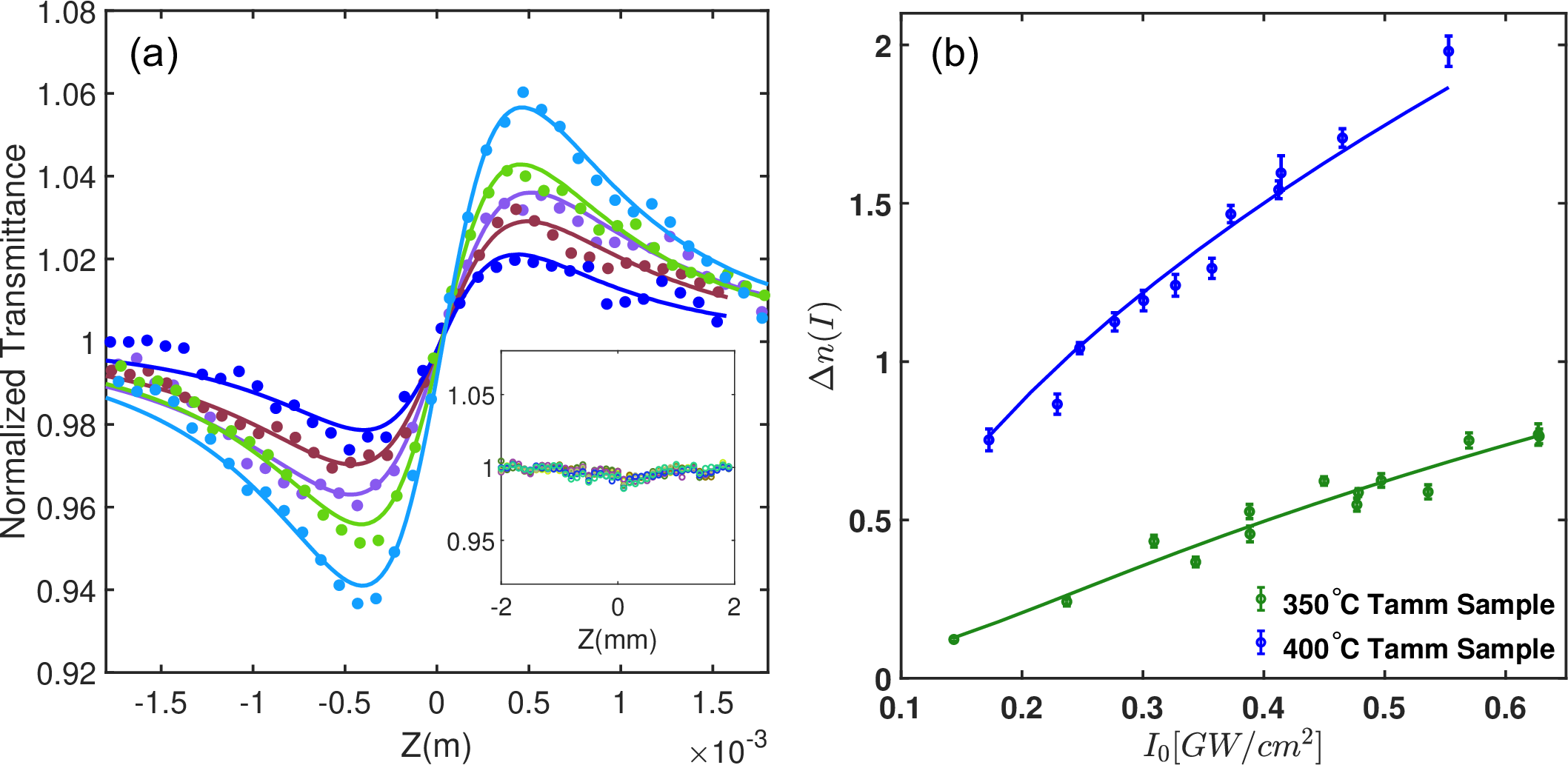}
  \caption{Z-scan characterization of the fabricated nonlinear Tamm state structure. (a) Representative closed-aperture scans measured at different pump intensities (increasing from 0.2 to 0.55 GW/cm$^{2}$) of the device annealed at $400^\circ$C. The continuous lines are the best fits obtained as discussed in the main text at a pump wavelength of 1250\,nm. The inset shows representative open-aperture scans of the same device. (b) Plot of $\Delta n$ as a function of the incident peak intensity of the device annealed at two temperatures. The solid lines represent the best fits of using the non-perturbative model from which the value of $\chi ^{(3)}_1$ is obtained.}
  \label{fig:1DTamm_zscan}
\end{figure}

The nonlinear optical response of the fabricated TPP structures were investigated using the Z-scan technique within the same setup discussed in section 2.3. Z-scan measurements at normal incidence were performed on the TPP structures annealed at $400^\circ$C and $350^\circ$C and excited at the resonant pump wavelength of $\lambda = 1250$nm. We also performed measurements on a reference TPP structure without the embedded ITO nanolayer. Open-aperture scans on the reference sample showed a weakly enhanced absorption of $\approx 4\%$ while closed-aperture scans were not detectable. Open-aperture Z-scan measurements on both the nonlinear TPP structures showed no measurable enhanced or saturable absorption, as displayed in the inset of figure \ref{fig:1DTamm_zscan}(a). A background subtraction was found to be necessary for the sample annealed at $350^\circ$C due to sample inhomogeneity. Here, a Z-scan is performed at low irradiance where nonlinear effects are not observed and the normalized Z-scan trace is subtracted from subsequent Z-scans at higher powers \cite{Z-scanOG}. No such background subtraction was necessary for the sample annealed at $400^\circ$C. Closed-aperture scans performed at different pump powers along with the fits for the $400^\circ$C-annealed structure are shown in figure \ref{fig:1DTamm_zscan}(a). These scans demonstrate remarkable changes in the refractive index of the 30\,nm ITO nanolayer enhanced by the coupling with the excited TPP state at the ENZ wavelength. In addition, the symmetric peak and valleys indicate the absence of nonlinear absorption effects in this structure, consistently with the open-aperture data. The refractive index nonlinear variation as a function of pump power for each TPP sample is reported in figure \ref{fig:1DTamm_zscan}(b) along with best fit using Eq. \ref{BH2}. The data indicate a dramatic enhancement of the non-perturbative Kerr nonlinearity of ITO nanolayers with a measured refractive index change $\Delta{n}\approx$ 2 in the TPP structure annealed at $400^\circ$C. As a comparison, a bare 30\,nm-thick nanolayer of ITO was also investigated and showed no measurable nonlinear signal in either open-or closed-aperture Z-scans even at the highest powers used in our setup. Calculations based on the TMM theory show that the field intensity delivered on the ITO nanolayer in the TPP photonic structure is $\approx$ 8$\times$ larger than the value achieved on a bare 30\,nm nanolayer. To the best of our knowledge, our findings demonstrate record-high nonlinear index change in 30 nm-thick ITO nanolayers due to the coupling of TPP surface states at ENZ wavelengths of ITO in a compact photonic structure. 

\section{Conclusions}
In summary, we studied the linear and nonlinear optical properties of magnetron sputtered ITO thin films and resonant photonic structures with ITO nanolayers coupled to TPP states. The linear optical properties of ITO thin films were characterized using broad band spectroscopic ellipsometry in conjunction with normal incidence transmission measurements. Our findings show that ITO thin films are accurately characterized by a TDL oscillator model over a wide spectral range. We employ post-deposition vacuum annealing of ITO thin films at temperatures ranging from 350$^\circ$-550$^\circ$C and demonstrate enhancement of optical nonlinearity with a maximum third-order susceptibility of $\chi^{(3)}_1 = 7.58\times10^{-17}$ $(m^2/V^2)$ at 400$^\circ$C annealing temperature. Furthermore, we show distinctive saturation effects in the intensity dependent nonlinear refractive index and extinction in excellent agreement with the general non-perturbative description of optical nonlinearity. We then fabricated a nonlinear Tamm state structure based on SiO$_2$/SiN dielectric DBR with an embedded ITO nanolayer. The parabolic mode dispersion of the fabricated TPP structures was characterized with normal incidence transmission and variable angle reflection measurements in excellent agreement with TMM theory. We demonstrated that TPP structures feature surface modes excited through free space at all investigated angles and polarizations with average optical field enhancement up to 5$\times$ inside the nonlinear ITO nanolayer. We further investigated the nonlinear optical response of the TPP structures via Z-scan technique. Our findings demonstrate non-perturbative refractive index variations as large as $\Delta n\approx 2$ driven by the near-field enhancement within ENZ nanolayers through strongly confined Tamm surface polariton states. This work serves as a stepping stone for the engineering of more efficient nonlinear devices and nanostructures with exceptional nonlinearity for applications to integrated all-optical data processing, spectroscopy, sensing and novel infrared photodetection modalities.


\section{Experimental Section}
\threesubsection{Thin Film Growth of Indium Tin Oxide}\\
ITO thin films were grown atop fused silica substrates via room-temperature radio-frequency magnetron sputtering using a Denton Discovery 18 sputtering system with a base pressure of $2\times10^{-7}$ Torr. We used a 99.99\% purity 3-in. ITO (In$_2$O$_3$/SnO$_2$ 90/10 wt.\%) target and sputtered the thin films in a pure argon environment at 5mTorr and 150W of RF power resulting in a deposition rate of 15.5 nm/min. All substrates were cleaned in piranha solution and plasma ashed in an oxygen environment prior to deposition. Post-deposition vacuum annealing was performed in a Mellen split tube furnace at a pressure of 3 mTorr. After deposition, four annealing temperatures ranging from 350 to 550 $^\circ$C were applied for one hour with a 20 minute ramp up/down time.
\newline
\threesubsection{Z-scan Experimental Setup}
ITO thin films were excited using a Spectra-Physics Inspire optical parametric oscillator (OPO) pumped by an 80MHz repetition rate Mai Tai HP Ti:sapphire tunable laser with 150 fs pulses with TEM00 idler output at near-infrared wavelengths. Two converging lenses (L1 and L2) act as a telescope that both magnifies the idler output and focuses the beam to a small spot size at the plane of a mechanical chopper. A reference beam is analyzed with a beam splitter (BS1) and the beam is focused onto the sample by a 30 mm focal length plano–convex lens(L3) down to a Gaussian beam waist value of $w_0 = 11 \mu m$ at the wavelength $\lambda = 1200\,nm$. The sample is mounted normal to the z-axis atop a computer-controlled motorized linear translation stage. The signal beam is split into two closed and open-aperture paths by a pellicle beam splitter (BS2) with the closed-aperture beam sampled in the far-field approximately 1 m away through a 200$\mu m$ pinhole and the entire open-aperture beam collected by a lens (L5). We attenuate the laser power by rotating a half-wave plate (HWP) positioned before a linear polarizer (LP) oriented for TM polarization. A long pass filter (LPF) is used directly after the OPO in order to extinguish the remaining second harmonic beam within the OPO cavity. The reference and the signals are all measured by a Tektronix 11000-series oscilloscope averaging over 64 acquisitions per z-translation interval. Consequently, we record the time evolution of Z-scan traces starting from the chopper opening rise time through the duration of its duty cycle.

\threesubsection{Fabrication of distributed Bragg Reflector}\\
A distributed Bragg mirror with 13 layers was fabricated by growing alternating layers of silicon nitride and silicon dioxide atop fused silica substrates via RF sputtering using a Denton Discovery 18 sputtering system with a base pressure of $2\times10^{-7}$ Torr. Silicon nitride films were reactively grown in an argon-nitrogen (2:1) environment with a 3-inch silicon target at a deposition pressure of 2.5\,mTorr and 200\,W of RF power resulting in a deposition rate of $\approx 5$nm/min. The silicon dixode thin films were grown in a pure Argon environment with a 3-inch silicon dioxide target at 5mTorr and 300W of RF power resulting in a deposition rate of 11\,nm/min.
Both of these materials were grown with a substrate temperature at $400^{\circ}$C during deposition. The optical constants and thickness of the grown silicon nitride and silicon dioxide thin films were characterized by variable-angle ellipsometry as discussed in the main text and in the supplementary material. The ellipsometry data, fits and extracted optical constants are reported in the supplementary section.

\medskip
\textbf{Supporting Information} \par 
Supporting Information is available from the Wiley Online Library or from the author.

\medskip
\textbf{Acknowledgements} \par 
This research was sponsored by the U.S. Army Research Office and accomplished under award number\newline W911NF2210110. We acknowledge Ayman A. Abdelhakeem who helped with the deposition of a test Bragg mirror during the early stages of this project and Dr. Fabrizio Sgrignuoli who helped us to develop the data fitting procedure for the Z-scan measurements.

\medskip

%
\bibliographystyle{MSP}
\bibliography{AOM_Bib}


\begin{figure}
  \includegraphics[width=\linewidth]{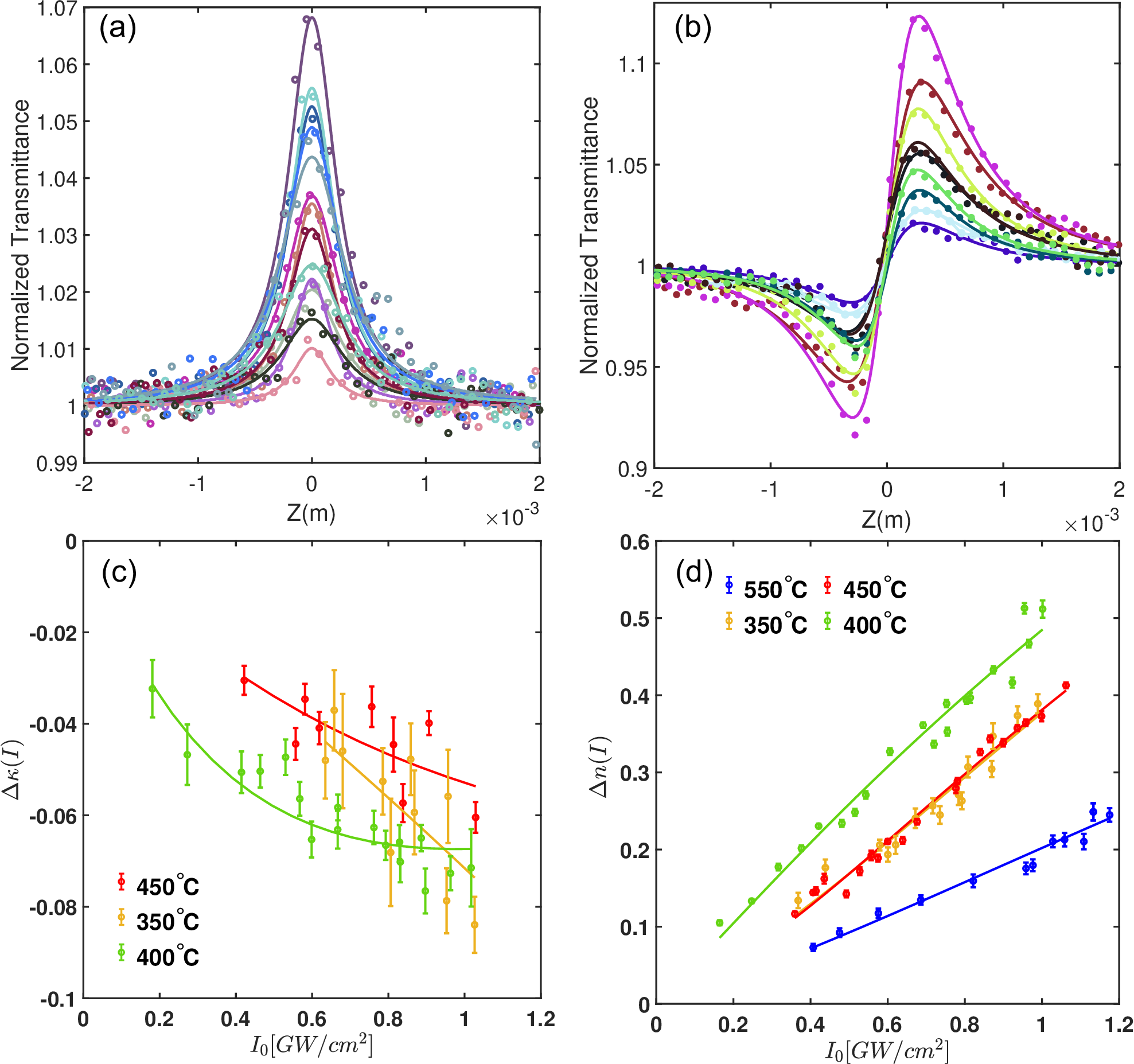}
  \caption{Z-scan nonlinear measurements of 300\,nm ITO thin films annealed at different temperatures. Representative (a) open- and (b) closed-aperture transmission curves for the sample annealed at $400^\circ$C measured at different pump intensities increasing from 0.2 to 1 GW/cm$^{2}$. (c-d) Plots of $\Delta k$ and $\Delta n$ as a function of the peak intensity extracted from open- and closed-aperture Z-scans, respectively. The solid lines represent fits of the data at $\lambda$=1200nm using the non-perturbative model discussed in the main text.}
  \label{fig:ITO z-scan_results}
\end{figure}

\end{document}